\documentclass[a4paper]{jpconf}
\usepackage{graphicx}
\begin{document}
\title{Recent results from NA61/SHINE and NA49}

\author{Anar Rustamov for the NA61/SHINE and NA49 Collaborations}

\address{Institut f\"ur Kernphysik, Goethe-Universit\"at Frankfurt, Frankfurt am Main, Germany}

\ead{a.rustamov@cern.ch}

\begin{abstract}
 Preliminary results on particle spectra and fluctuations  in inelastic p+p interactions measured by NA61/SHINE at the CERN SPS 
 are presented and its future ion program is discussed. We also show results on the excitation functions of event-by-event 
 particle ratio fluctuations in central Pb+Pb collisions
 from NA49 using a novel approach. We found a dependence of the results on phase space coverage  which appears to explain the reported difference between measurements of NA49 and those of STAR in central Au+Au collisions.
\end{abstract}

\section{Introduction}
As the fundamental theory of strong interactions, Quantum Chromodynamics (QCD), is asymptotically free, the created matter at high temperature and/or high density may be dominated by the state of quasi-free quarks and gluons referred to as the Quark-Gluon Plasma (QGP). By colliding heavy-ions at high energies one hopes to heat and/or compress the matter to energy densities at which the production of the QGP begins. Indeed, lattice QCD calculations predict this new phase at high temperatures. The motivation of investigating relativistic heavy ion collisions is the experimental study of hadronic matter under extreme conditions. A wealth of ideas in the past few decades have been proposed to explore the phase structure of strongly interacting matter. Among other possible probes event-by-event fluctuations of different observables may be sensitive  to the transitions between hadronic and partonic phases. In particular, the location of the critical point may be signalled by a characteristic pattern in the energy and system size dependence of the measured fluctuation signals. 
	
The SPS Heavy Ion and Neutrino Experiment (NA61/SHINE) is operating since 2007 in the North Area of the 
CERN Super Proton Synchrotron (SPS). NA61/SHINE uses the apparatus of the NA49~\cite{NA49_NIM} experiment 
with numerous upgrades~\cite{NA61_SPSC2006}. Its large acceptance and excellent particle identification 
capabilities~\cite{NA49_NIM, NA61_SPSC2006} make possible a rich physics program, in particular 
the ion program which is the main subject of this contribution. 
The ultimate goal is to explore the phase diagram of strongly interacting matter using collisions 
of different systems from p+p, through p+A, to A+A collisions at projectile momenta 
of 13\textit{A}, 20\textit{A}, 30\textit{A}, 40\textit{A}, 80\textit{A} and 158\textit{A} GeV/c. 
The energy scan of p+p reactions was completed 
in 2009-2011 and preliminary results will be presented. Data on Be+Be collisions at 
three top beam momenta and at 13\textit{A} GeV/c were recorded in 2011 and 2012. In 2013 Be+Be data at the remaining energies will be collected.
The energy scans of p+Pb, Ar+Ca and Xe+La collisions are foreseen until 2016. In addition a beam energy scan of Pb+Pb collisions is 
planned. The explored energy range probes an important region in the phase diagram of strongly interacting matter. 
Indeed, the NA49 collaboration reported signals for the onset of deconfinement encoded in non-monotonic 
behavior of excitation functions of several hadronic observables~\cite{NA49_horn}. These observations have been recently confirmed by the RHIC and LHC data ~\cite{NA49_Rustamov}. Moreover,  
the critical point at the location predicted by several theory groups~\cite{Zoltan} can be probed at SPS energies.
  
  \begin{figure}[htbp]
\begin{center}
\includegraphics[width=0.35\textwidth]{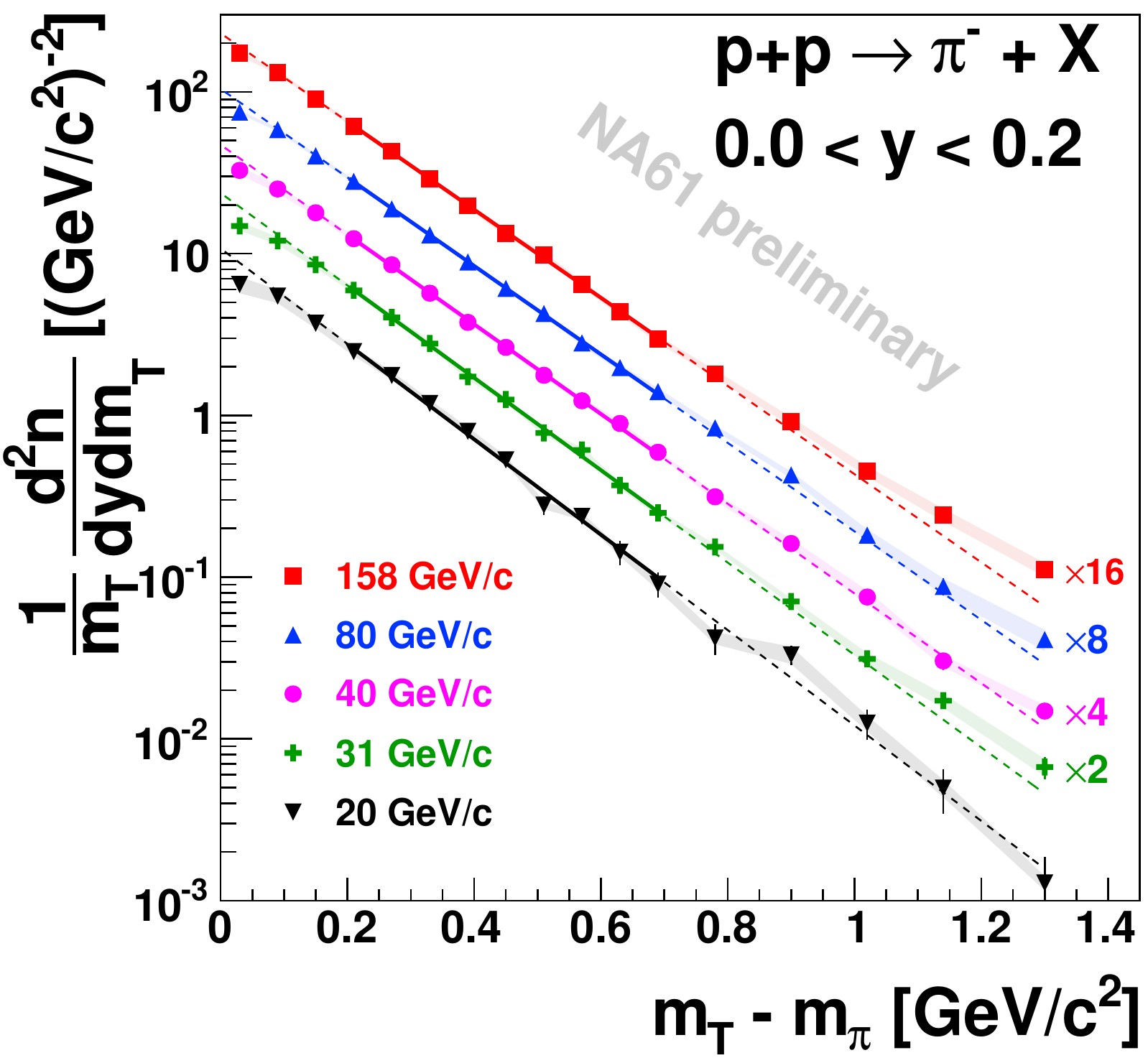}
\includegraphics[width=0.367\textwidth]{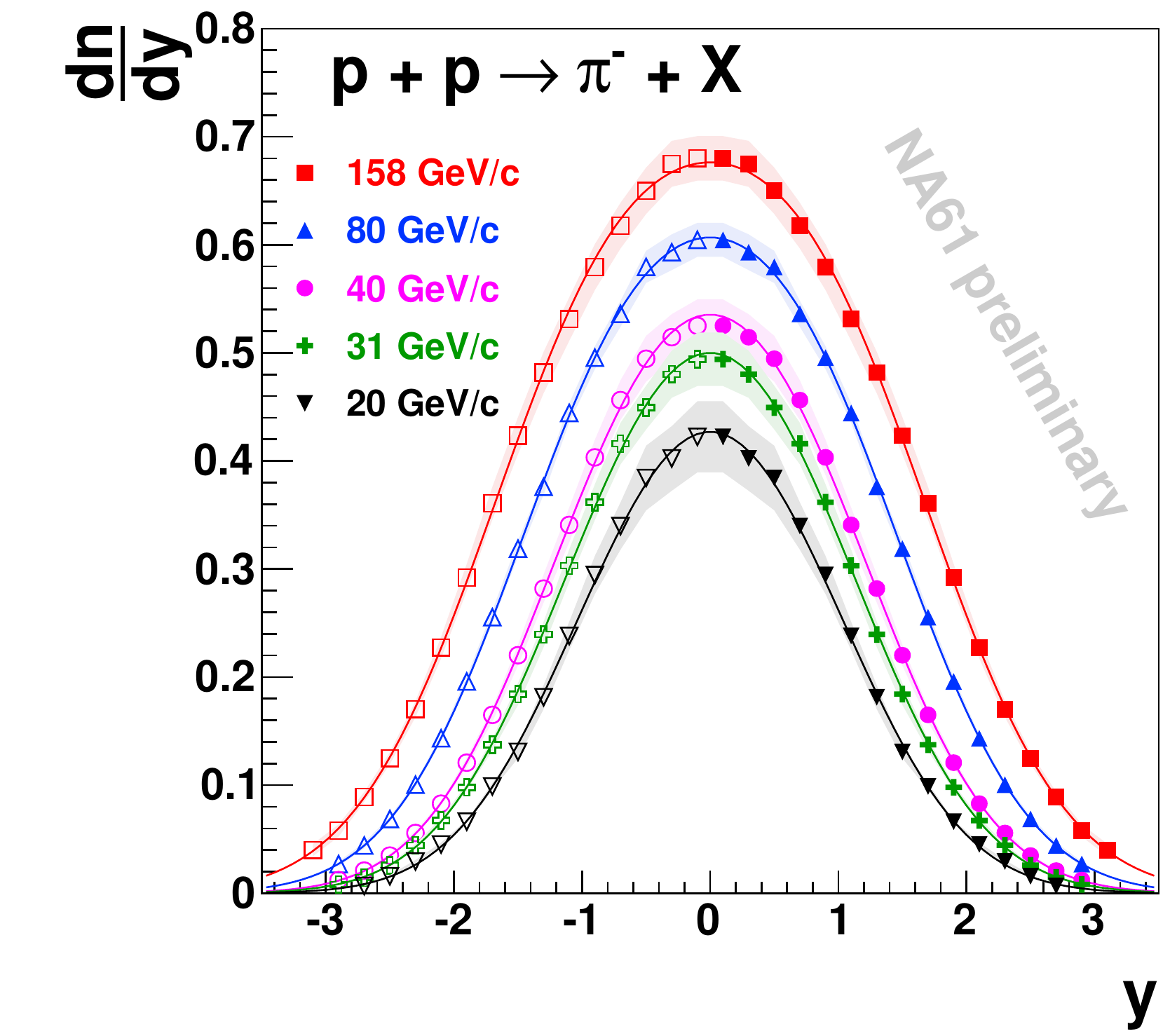}
\end{center}
\caption{(Color online) Left: Transverse-mass spectra at mid-rapidity for negatively charged pions fitted with an exponential function in the 0.2 $< m_{T} - m_{\pi} <$ 0.7 GeV/c$^{2}$ interval. Right: Rapidity spectra of negatively charged pions fitted with the sum of two Gaussian functions. Results refer to inelastic p+p interactions at beam momenta of 20-158 GeV/c.  }
\label{part_spectra}
\end{figure}
  
\section{Inclusive particle spectra in p+p interactions}
Results from the energy scan of p+p interactions will serve 
as an important reference for interpretation of the p+A and A+A data at corresponding beam energies collected with the same apparatus.
Figure~\ref{part_spectra} (left) presents the transverse mass ($m_T$) spectra at mid-rapidity for negatively 
charged pions in inelastic p+p collisions at 20, 31, 40, 80 and 158 GeV/c~\cite{NA61_SPSC2012}. 
The results are corrected for reconstruction inefficiencies, acceptance, and feed-down using Monte Carlo simulations. 
The $m_{T}$ spectra are extrapolated by fitting the distributions with an exponential function. Figure~\ref{part_spectra} (right)  
depicts the $m_{T}$ integrated rapidity spectra. The latter allow to obtain the mean number of negatively charged pions 
in full phase space from which one can calculate the total number of  pions using isospin coefficients. The NA61/SHINE pion multiplicities are in good agreement with the world data~\cite{NA61_SPSC2012}. 

\begin{figure}[htbp]
\begin{center}
\includegraphics[width=0.329\textwidth]{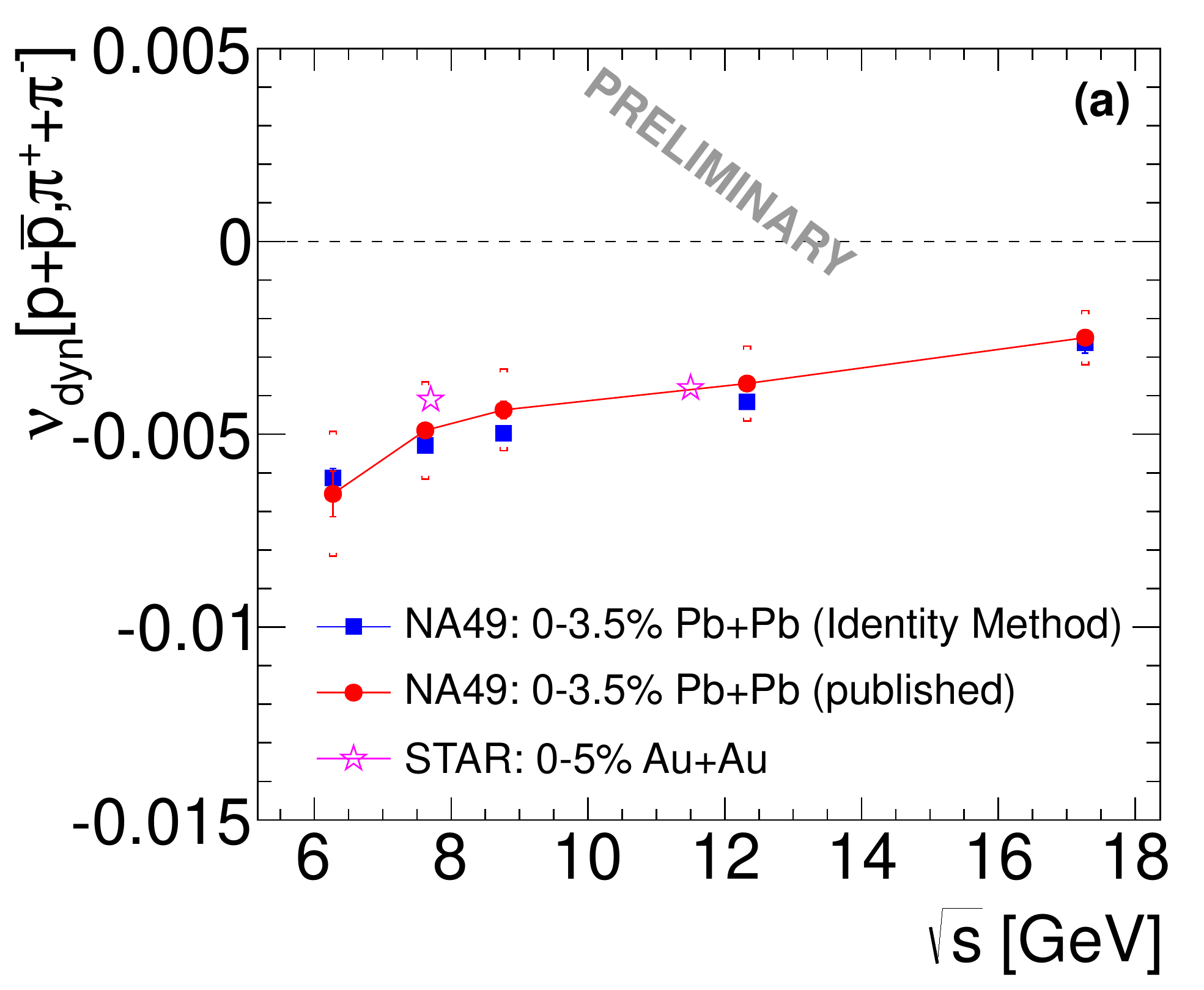}
\includegraphics[width=0.329\textwidth]{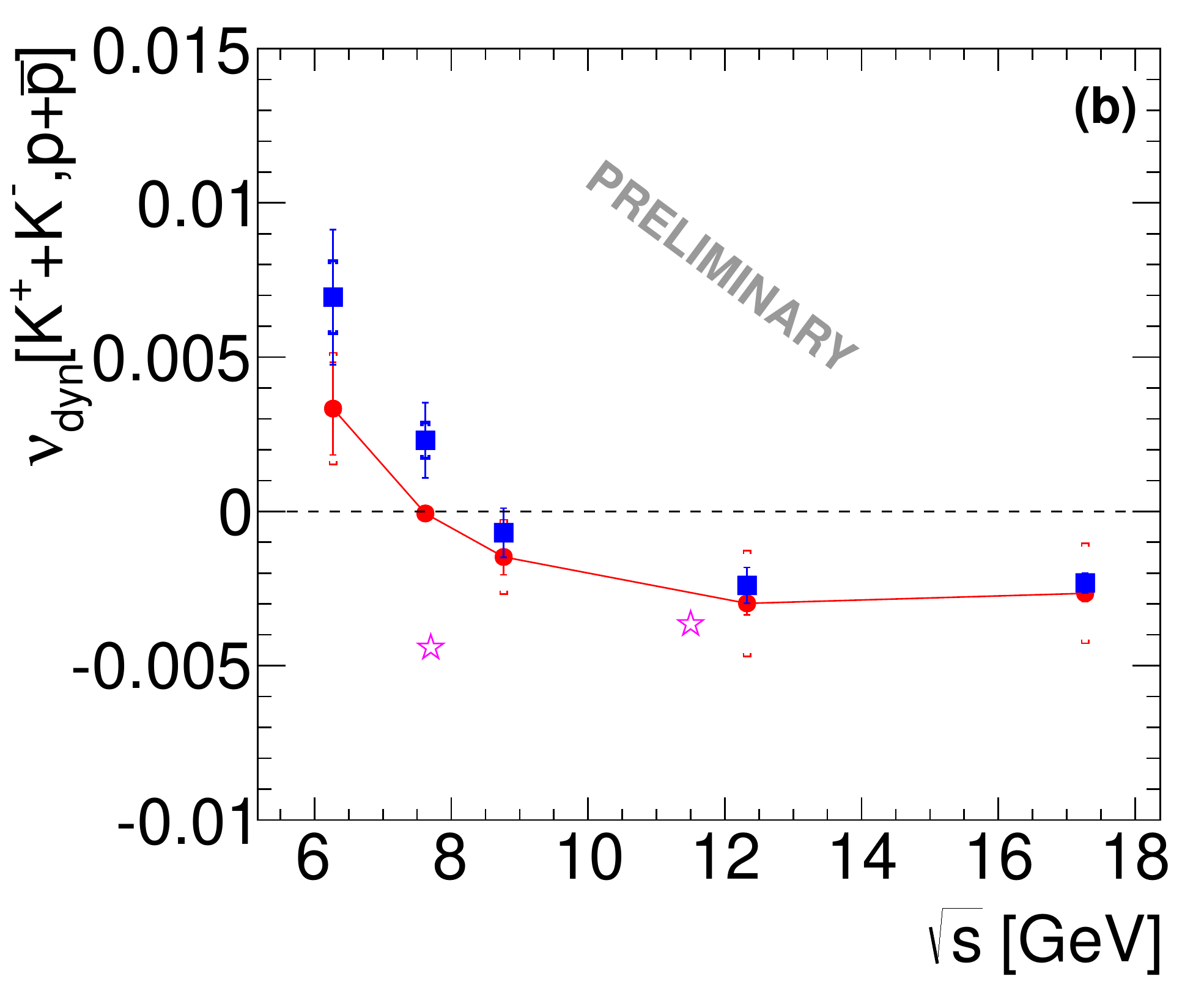}
\includegraphics[width=0.329\textwidth]{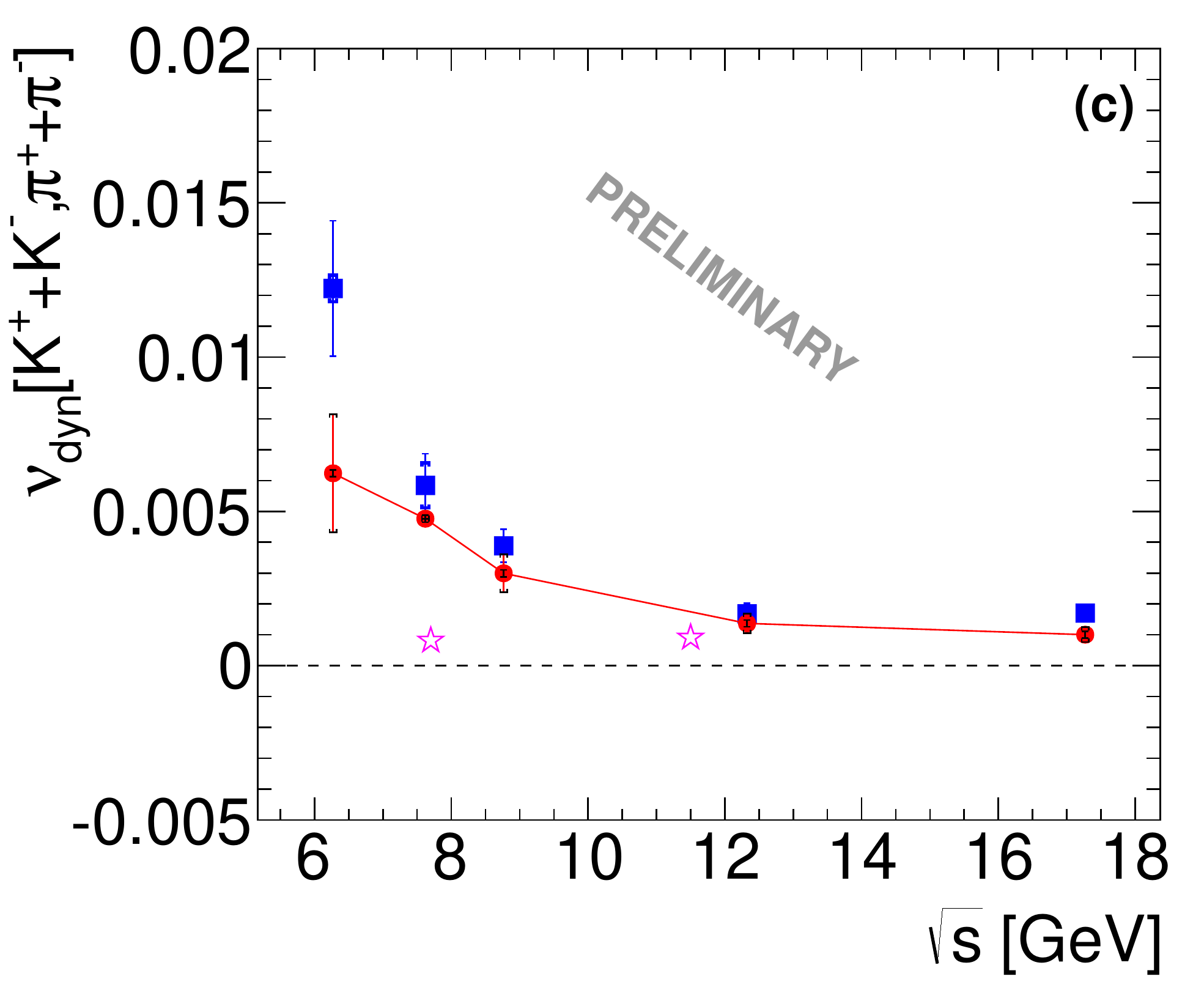}
\end{center}
\caption{(Color Online) Energy dependence of (a) $\nu_{dyn}[p+\bar{p},\pi^{+}+\pi^{-}]$, 
(b) $\nu_{dyn}[K^{+}+K^{-},p+\bar{p}]$ and (c) $\nu_{dyn}[K^{+}+K^{-},\pi^{+}+\pi^{-}]$. 
Results from the Identity Method for central Pb+Pb data of NA49 are shown by blue squares. 
Published NA49 results are indicated by circles. Stars represent results 
of the STAR collaboration for central Au+Au collisions. 
}
\label{nu_dyn}
\end{figure}

\section{Event by Event fluctuations in Pb+Pb and p+p collisions.}

A motivation for our work presented in this section is to shed light on reported differences between
the results from STAR and NA49 on fluctuations of identified hadrons~\cite{NA49_fluct1, NA49_fluct2, STAR_fluct}. The studied measure of dynamical particle 
ratio fluctuations $\nu_{dyn}[A,B]$ is defined as~\cite{voloshin_nu}:
\begin{equation}
\nu_{dyn}[A,B]=\frac{\left<A(A-1)\right>}{\left<A\right>^{2}} + \frac{\left<B(B-1)\right>}{\left<B\right>^{2}} - 2\frac{\left<AB\right>}{\left<A\right>\left<B\right>} , 
\end{equation}
where A and B stand for multiplicities of different particle species. 

\begin{figure}[htbp]
\begin{center}
\includegraphics[width=0.32\textwidth]{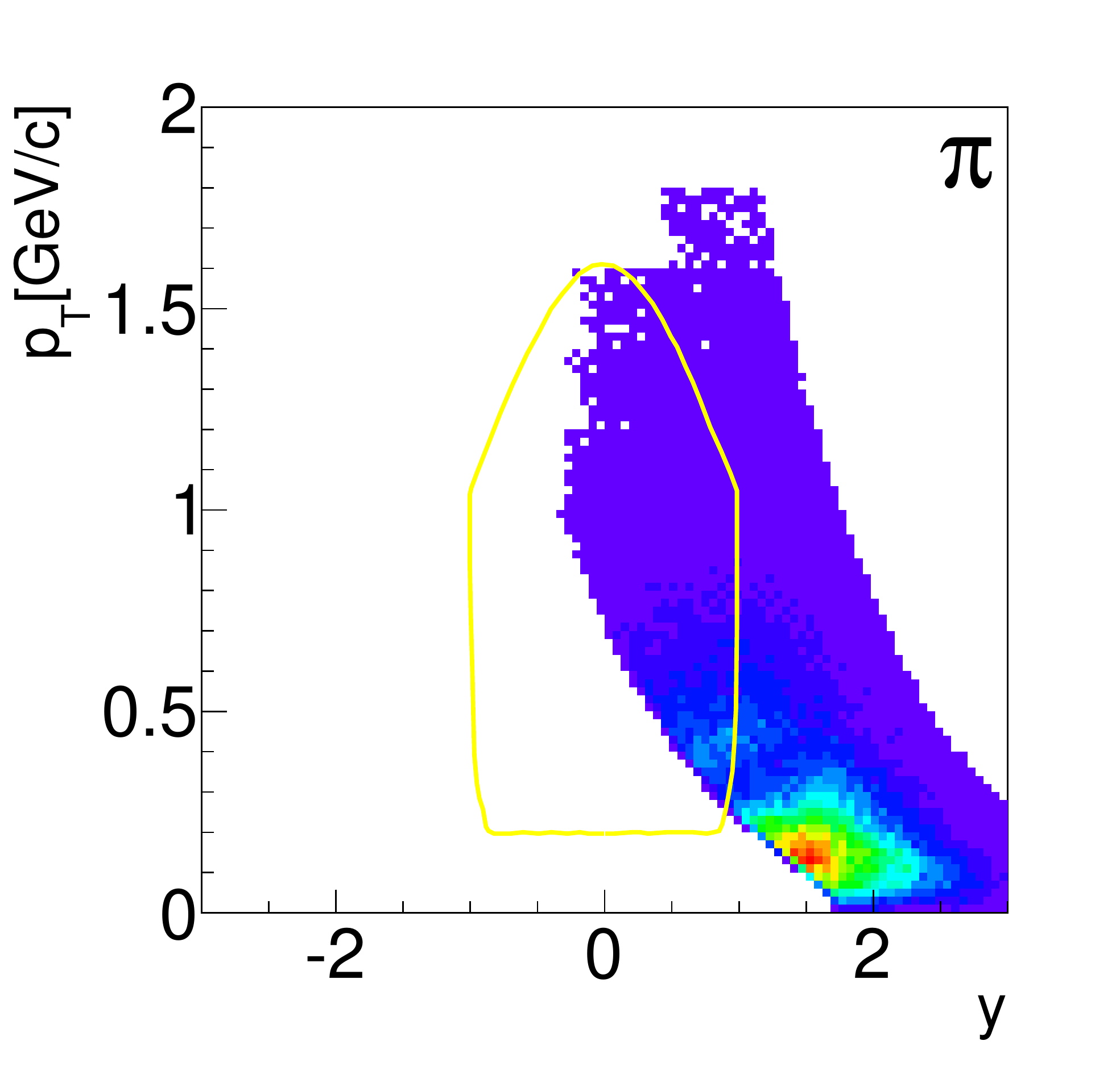}
\includegraphics[width=0.32\textwidth]{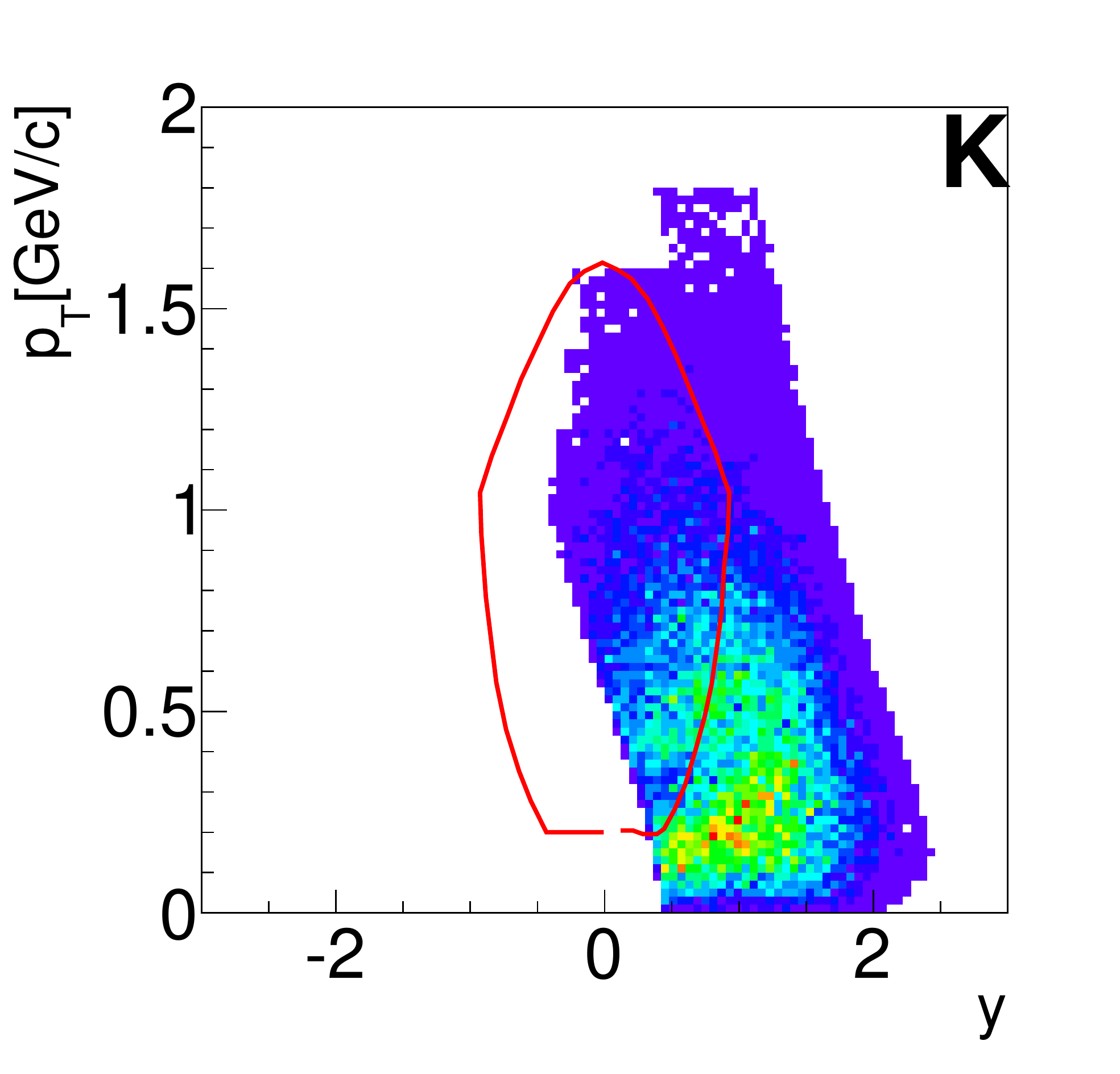}
\includegraphics[width=0.32\textwidth]{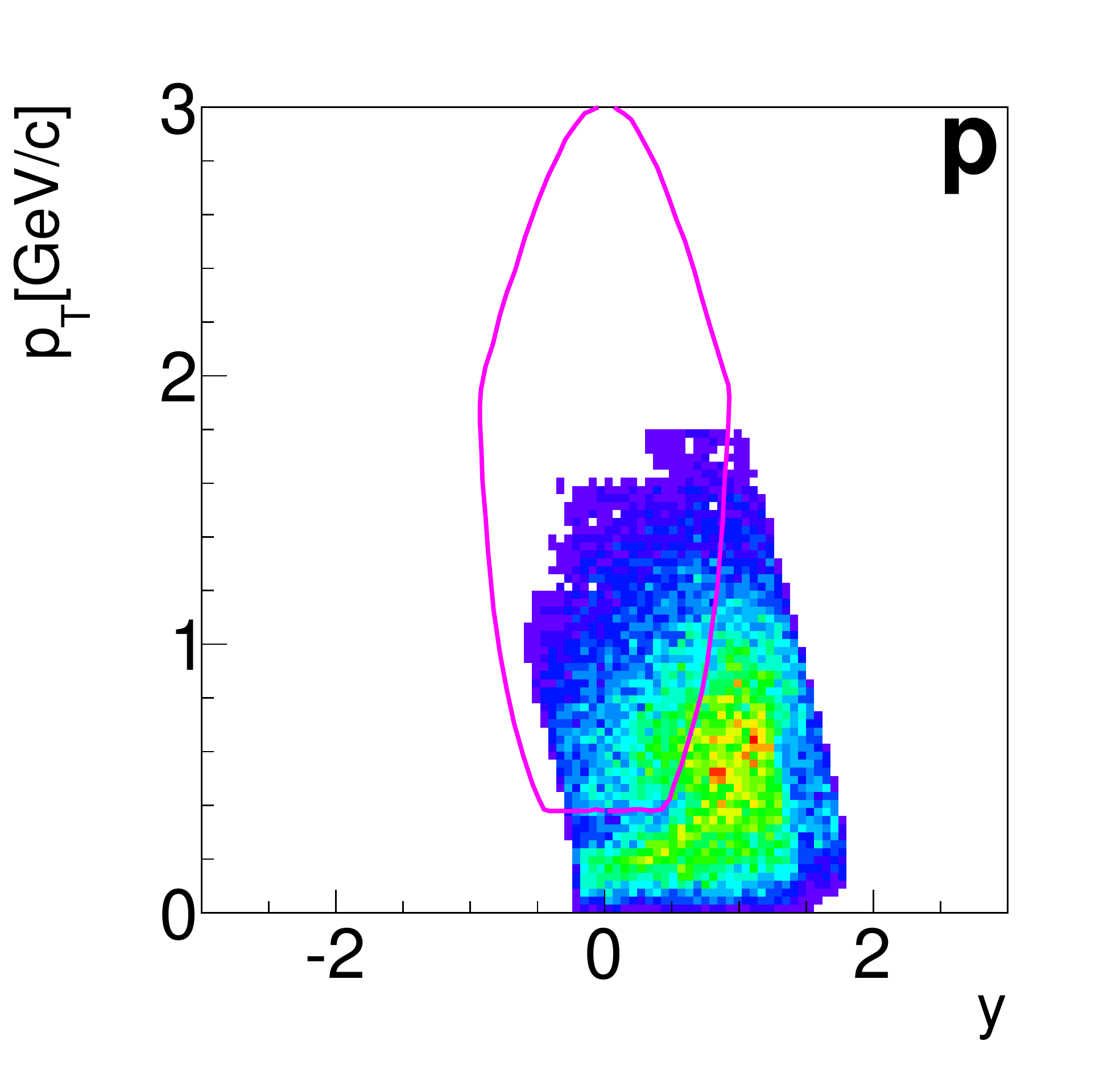}
\includegraphics[width=0.32\textwidth]{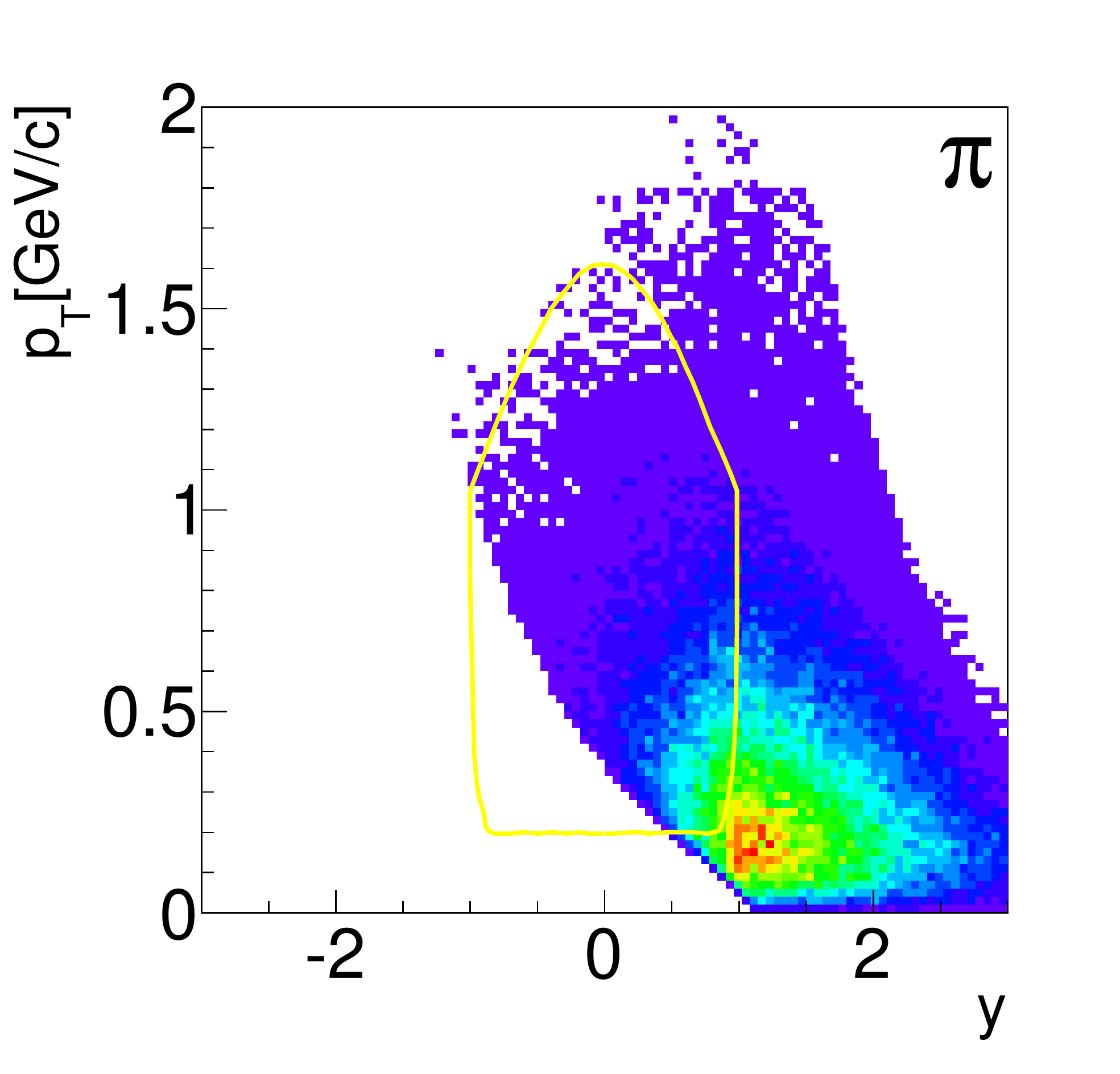}
\includegraphics[width=0.32\textwidth]{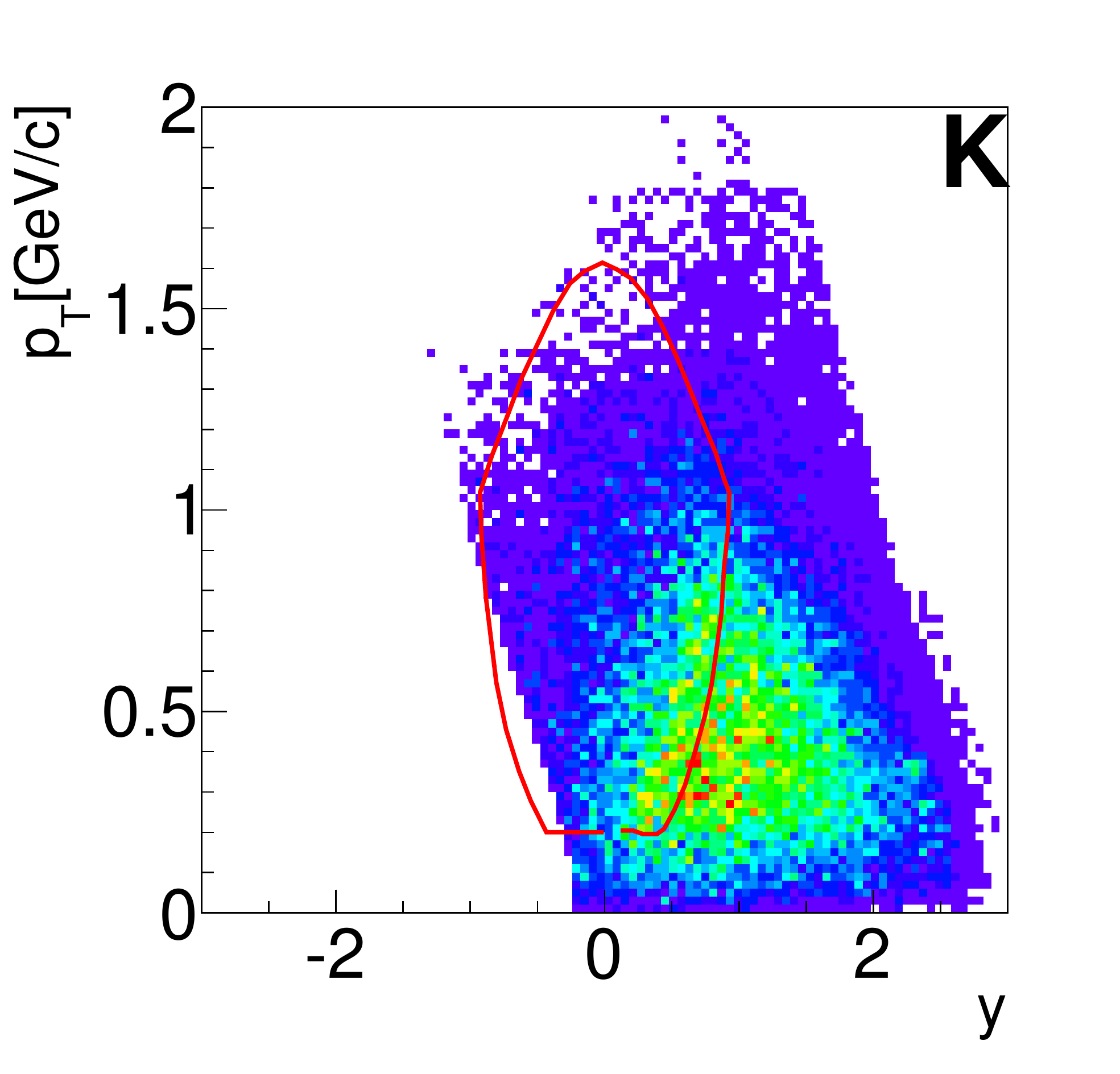}
\includegraphics[width=0.32\textwidth]{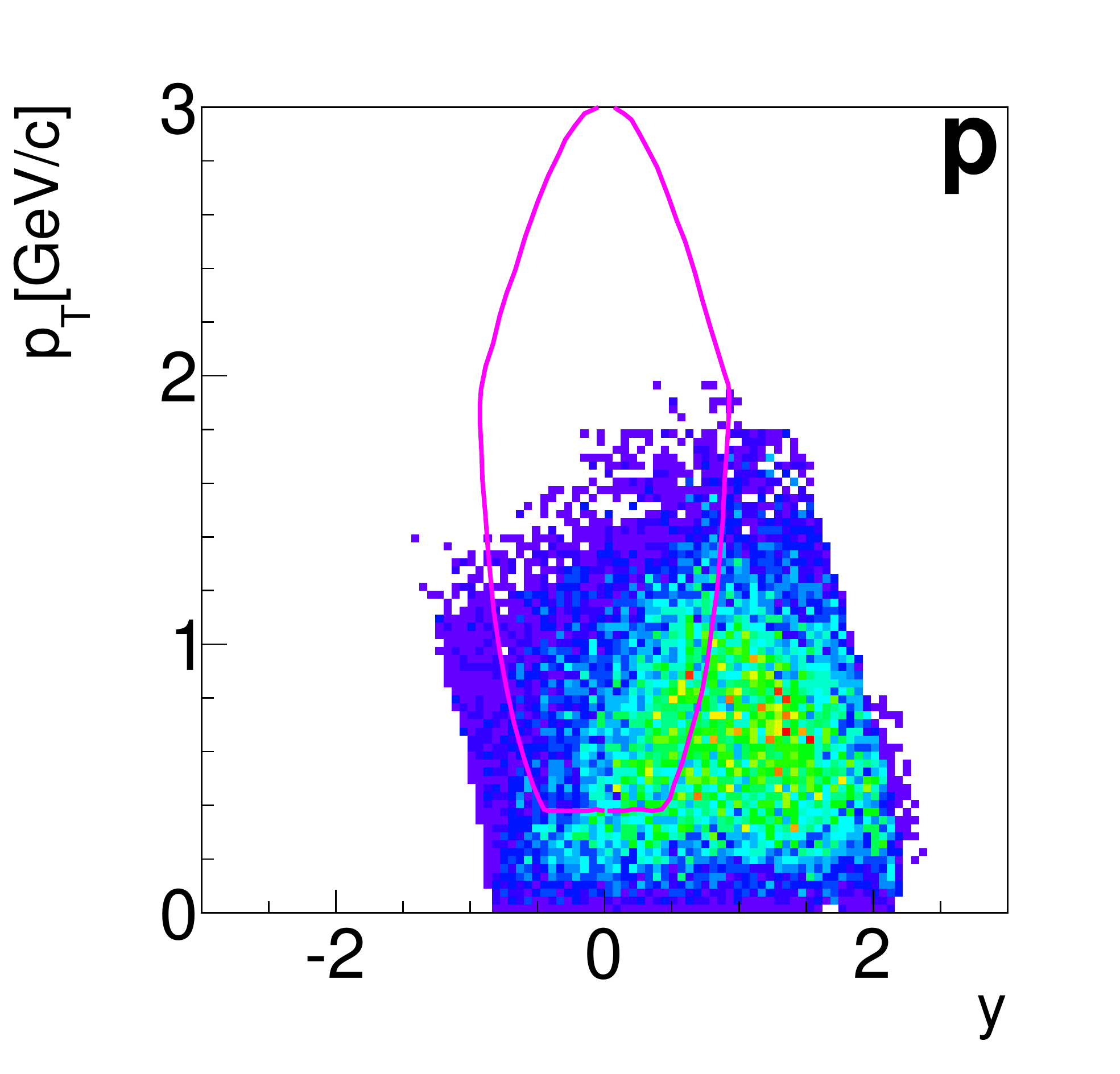}
\end{center}
\caption{(Color Online) The phase space coverage of pions, kaons and protons in the acceptance of the NA49 experiment for the Pb+Pb collisions at 30\textit{A} GeV/c (upper panel) and 158\textit{A} GeV/c (lower panel) used for the event-by-event fluctuations presented in Fig.~\ref{nu_dyn}. The acceptance of the STAR apparatus at corresponding center-of-mass energies are depicted with colored lines. 
}
\label{phase_space}
\end{figure}

As is evident from Eq.(1) all second moments of the multiplicity distributions of particles A and B have to be known
in order to construct $\nu_{dyn}$[A,B]. The standard approach is to identify these particles event-by-event. 
However, this approach is hampered by incomplete particle identification, which is taken care of by either selecting suitable phase space regions or by applying a fitting 
procedure event-by-event. The latter typically introduces artificial correlations which are usually corrected for 
by the event mixing technique. 

\begin{figure}[htbp]
\begin{center}
\includegraphics[width=0.329\textwidth]{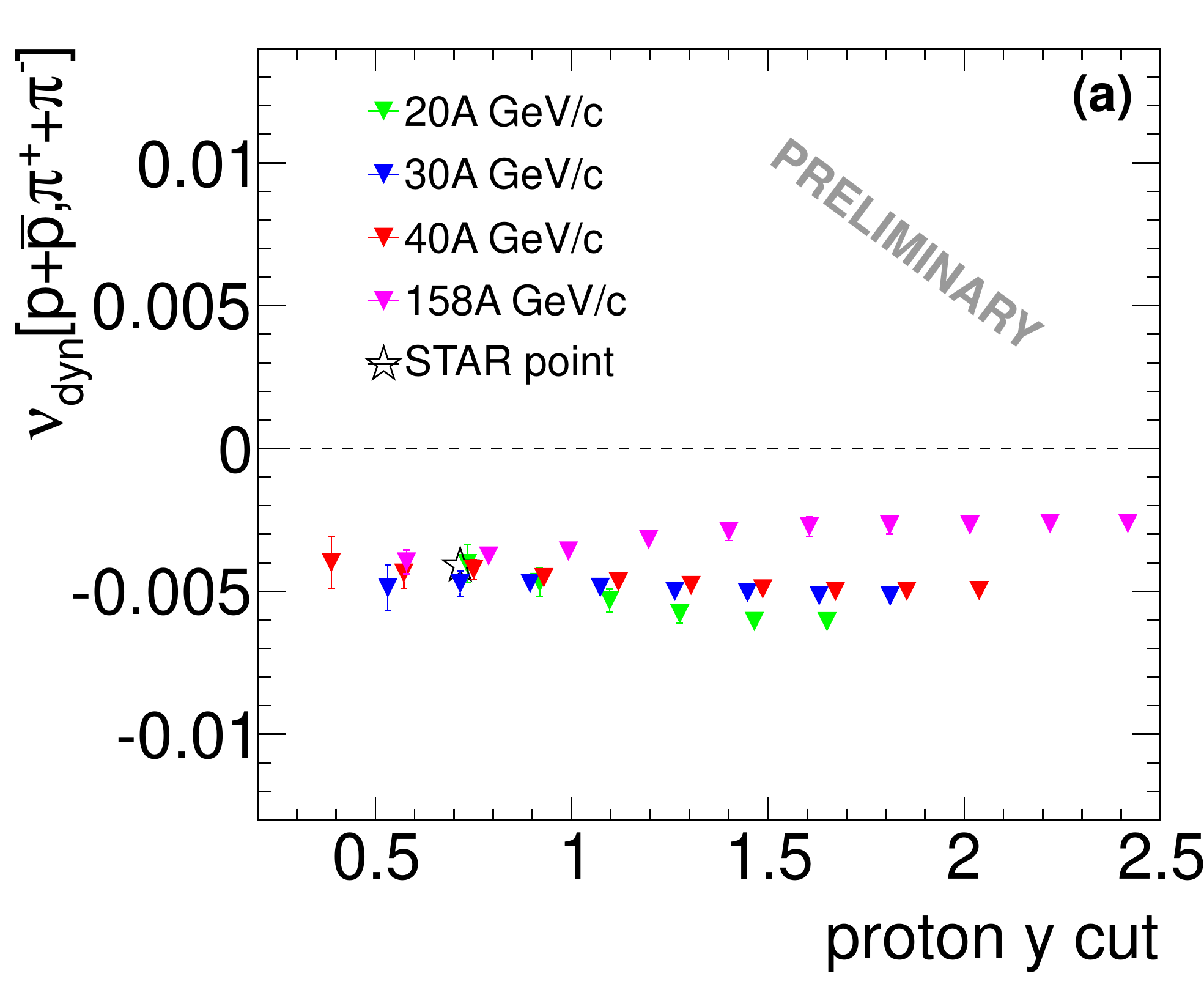}
\includegraphics[width=0.329\textwidth]{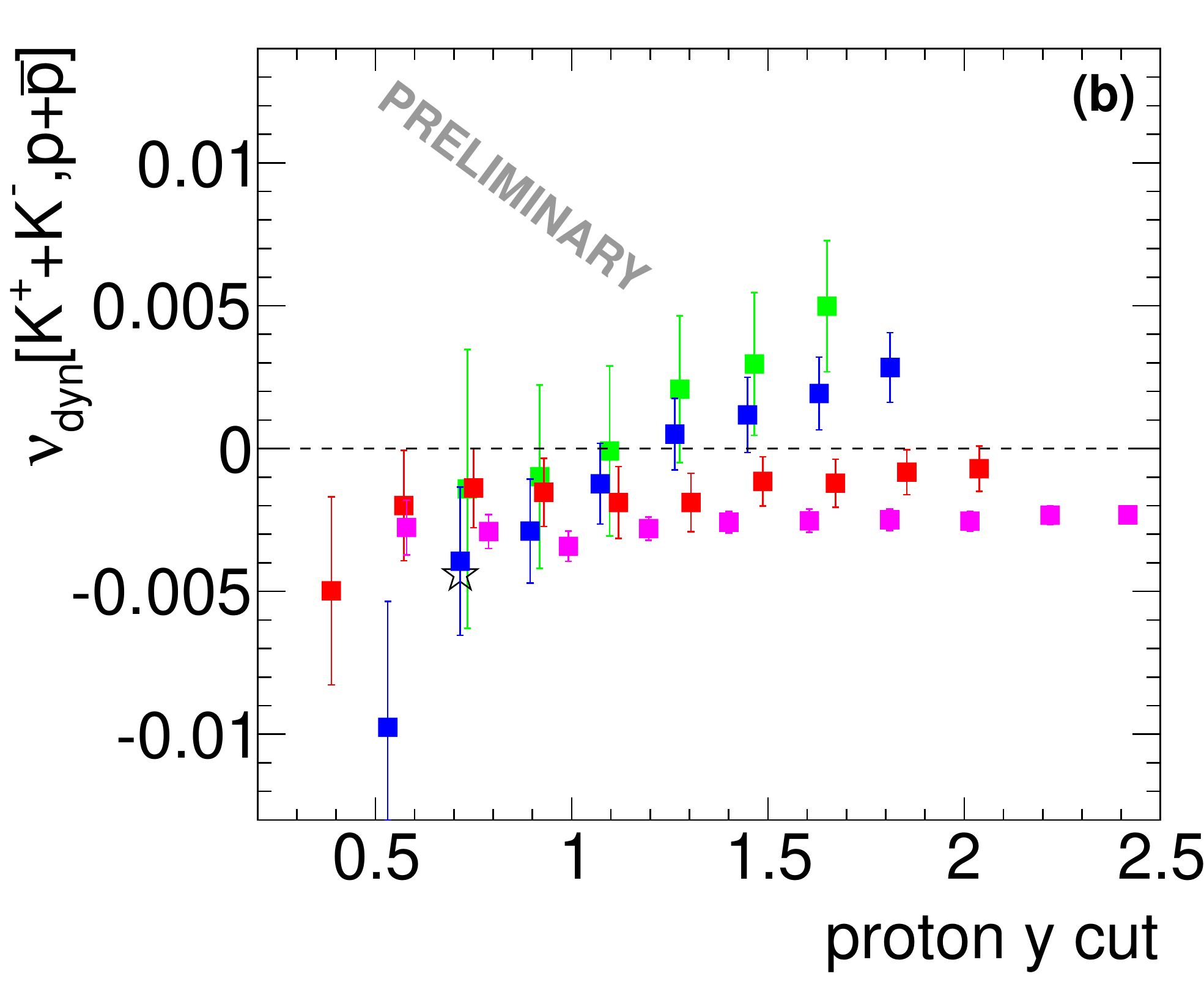}
\includegraphics[width=0.329\textwidth]{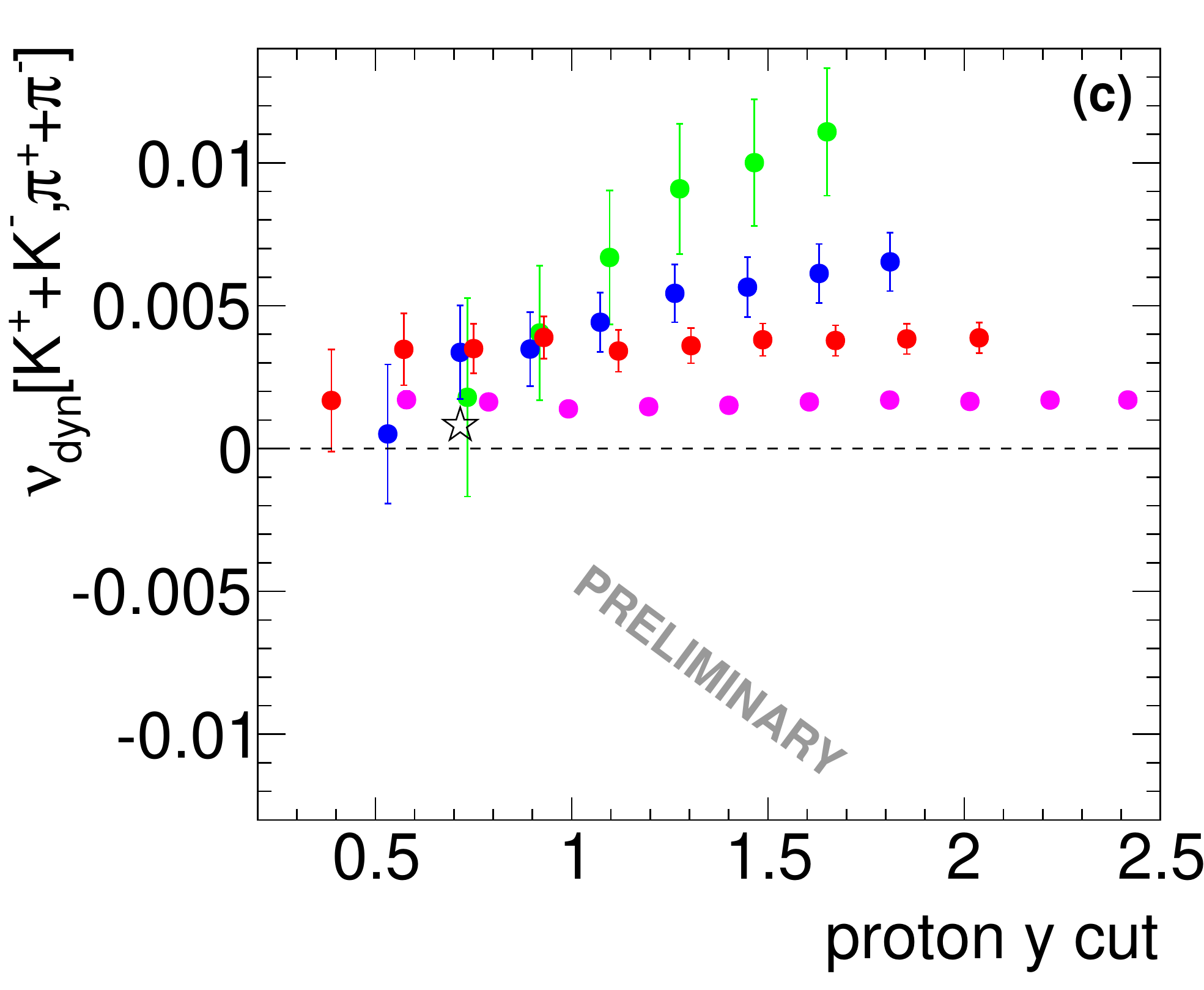}
\end{center}
\caption{(Color Online) Acceptance dependence of (a) $\nu_{dyn}[p+\bar{p},\pi^{+}+\pi^{-}]$, 
(b) $\nu_{dyn}[K^{+}+K^{-},p+\bar{p}]$ and (c) $\nu_{dyn}[K^{+}+K^{-},\pi^{+}+\pi^{-}]$ 
in central Pb+Pb collisions of NA49 (triangles, squares, dots).
Stars show measurements of the STAR collaboration.
Results are plotted versus the maximum of the proton rapidity at  $p_{T}$=0. 
}
\label{acc_dep}
\end{figure}

In this context a novel approach, called \emph{Identity Method}~\cite{identity1, identity3} was developed 
for fluctuation analysis. It uses a probabilistic approach which avoids the event-by-event fitting as well as has a rigorous mathematical derivation. Furthermore, there is no need for corrections based on event mixing anymore. 
Results using the \emph{Identity Method} are presented in Fig.~\ref{nu_dyn} for central Pb+Pb data of NA49 and 
compared to published NA49 and STAR results from its Beam Energy Scan (BES) program~\cite{STAR_fluct}. The energy dependence of $\nu_{dyn}[p+\bar{p},\pi^{+}+\pi^{-}]$ is 
consistent with the NA49 publication. Moreover the increasing trend of the excitation functions of  
$\nu_{dyn}[K^{+}+K^{-},p+\bar{p}]$ and $\nu_{dyn}[K^{+}+K^{-},\pi^{+}+\pi^{-}]$ is confirmed by this analysis. 
It was found in Ref.~\cite{Tim_Volker} that $\nu_{dyn}$  exhibits an intrinsic dependence on the multiplicities 
of accepted particles. 
Published NA49 results are essentially performed 
at forward rapidities, whereas the STAR acceptance covers the mid-rapidity region without the low $p_{T}$ range. As an example we illustrate in Fig.~\ref{phase_space} the phase space coverage of pions, kaons and protons at different projectile energies in the acceptance of NA49. In the same figure the acceptance of the STAR apparatus at corresponding center-of-mass energies are presented with colored lines. 
We, therefore, studied the acceptance 
dependence of $\nu_{dyn}$ by performing the analysis in different acceptance bins going from forward rapidity to 
mid-rapidity. Technically different acceptance bins were selected by applying several upper momentum cuts for each reconstructed track and calculating the corresponding maximum rapidity of protons at $p_{T}$=0. Thereafter we will call this quantity a proton rapidity cut. 
The dependence of obtained results on the proton rapidity cut is illustrated in Fig.~\ref{acc_dep}. At 20\textit{A} and 30\textit{A} GeV/c $\nu_{dyn}[K^{+}+K^{-},p+\bar{p}]$ and $\nu_{dyn}[K^{+}+K^{-},\pi^{+}+\pi^{-}]$
show a strong acceptance dependence, and eventually hit the STAR point in a particular acceptance bin. Interestingly
the acceptance dependence  weakens above 30\textit{A} GeV/c where no difference was observed with STAR.
It is also remarkable that $\nu_{dyn}[p+\bar{p},\pi^{+}+\pi^{-}]$ shows little acceptance dependence. 
This detailed study of fluctuation results in different acceptance bins appears to explain the 
difference between the STAR BES and NA49 measurements.


\begin{figure}[htbp]
\begin{center}
\includegraphics[width=0.329\textwidth]{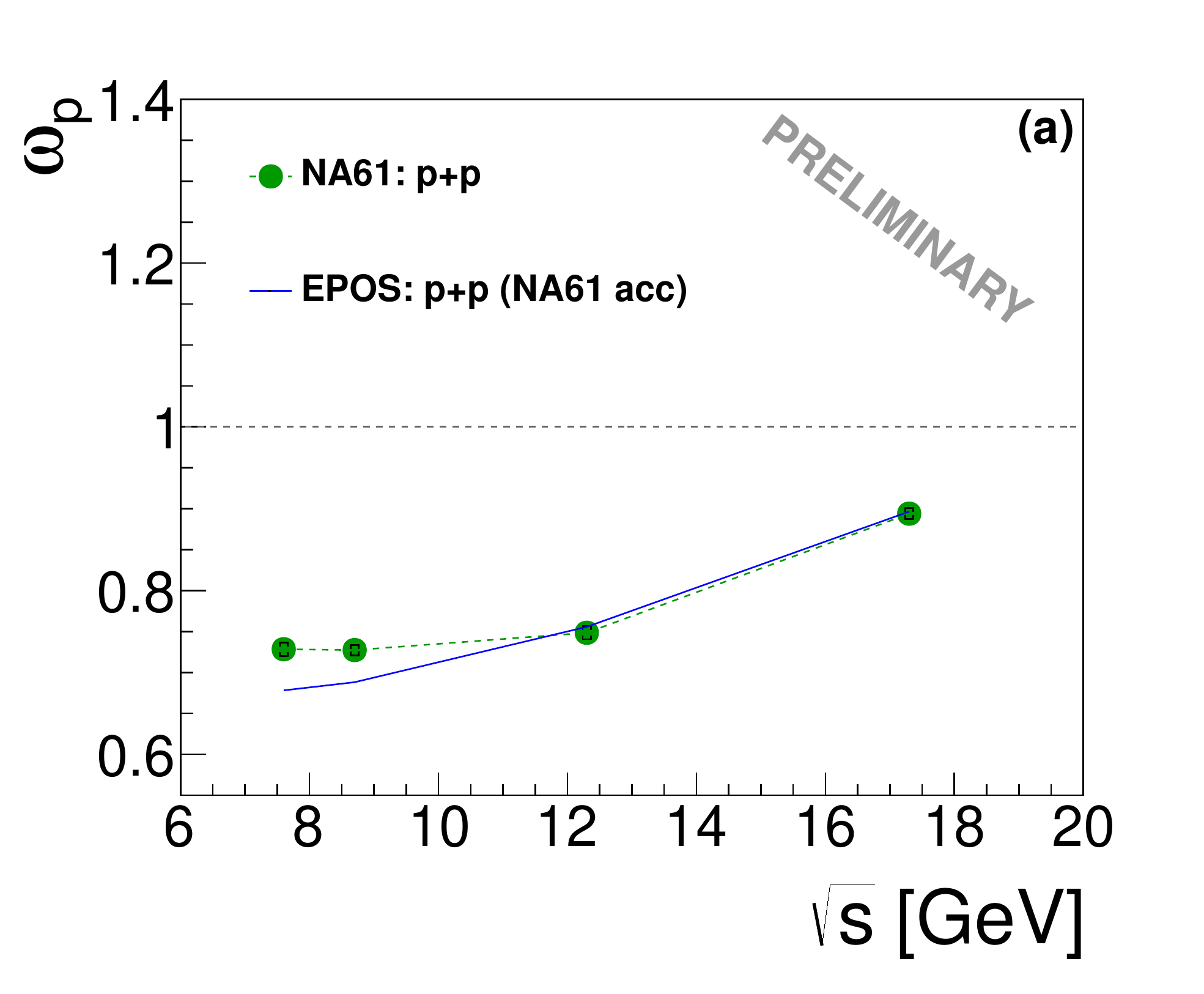}
\includegraphics[width=0.329\textwidth]{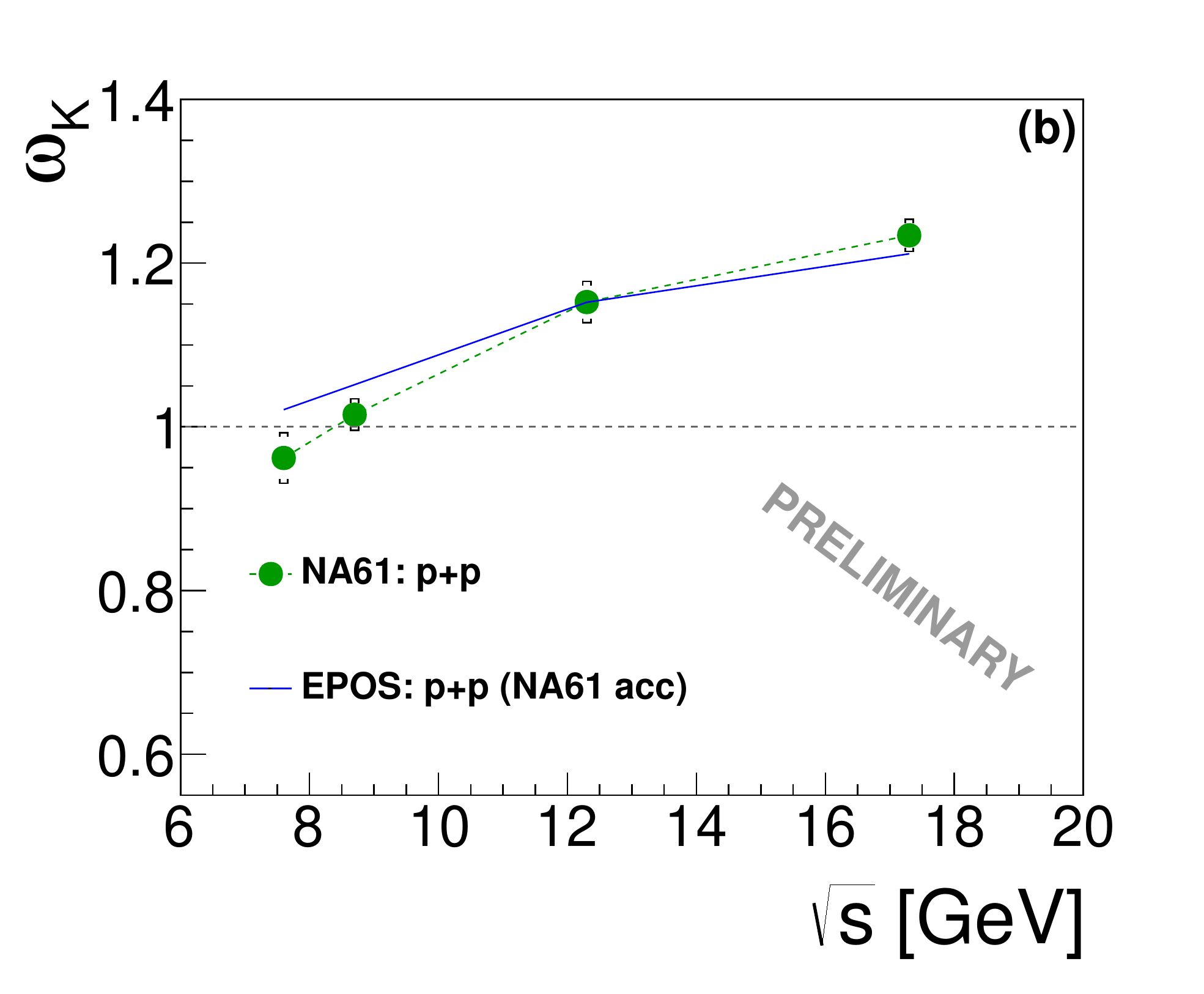}
\includegraphics[width=0.329\textwidth]{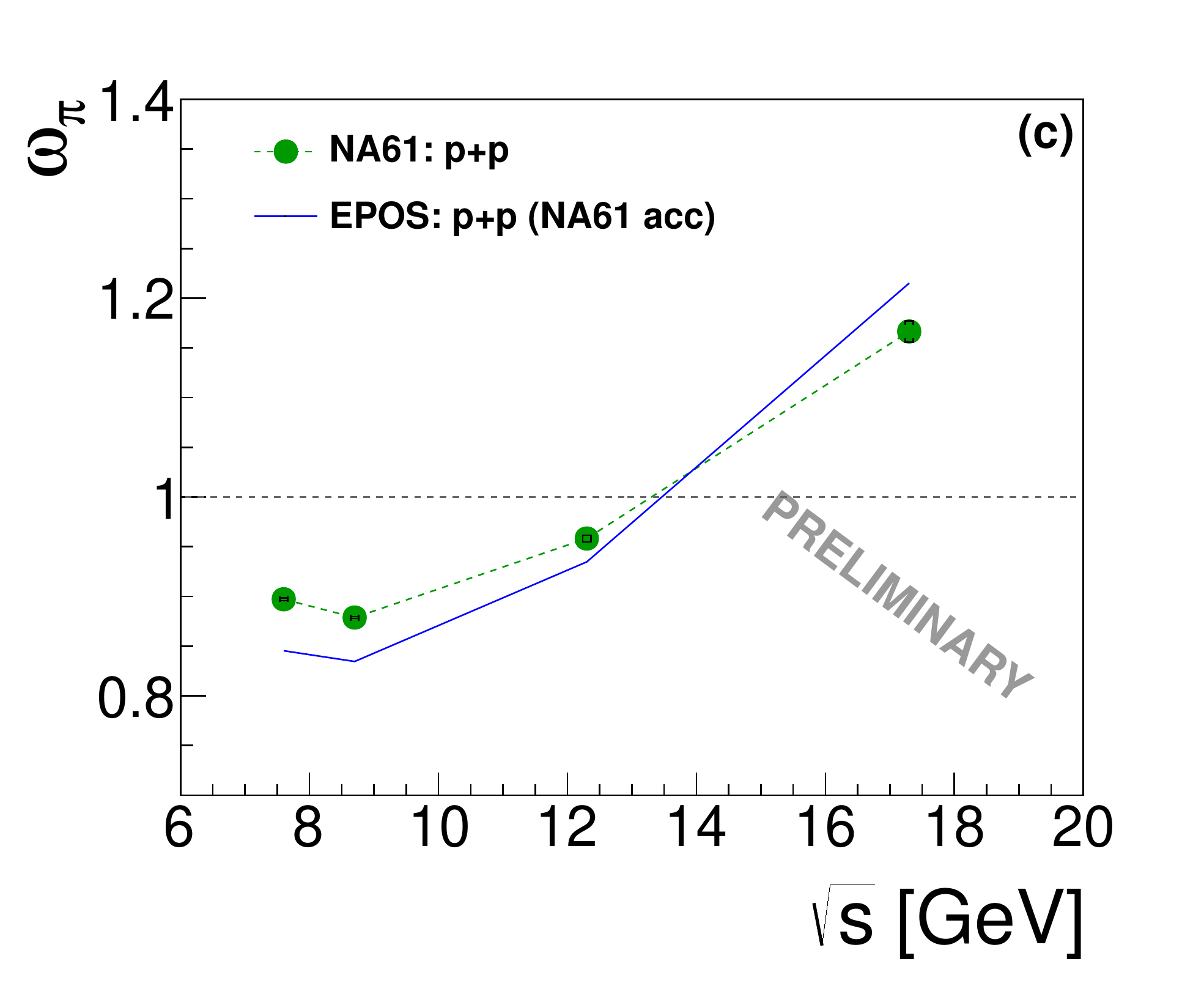}
\end{center}
\caption{(Color Online) Scaled variance for (a) $p+\bar{p}$, (b) $K^{+}+K^{-}$ and (c) $\pi^{+}+\pi^{-}$ 
multiplicity distributions for p+p collisions. The NA61/SHINE results are shown by filled circles and compared to 
EPOS model calculations depicted by a solid blue curve.}
\label{scaled_var}
\end{figure}

As mentioned in the introduction, the obtained fluctuation signals from A+A collisions should be compared systematically to the baseline obtained from analysis of p+p interactions at corresponding energies. It is however important to properly take into account trivial differences between A+A and p+p collisions e.g., in the size of the colliding systems. In thermodynamics the quantities which are proportional to the volume of the system are called extensive quantities. For example, the mean number of particles in a relativistic gas within the grand canonical ensemble is an extensive quantity. The ratio of two extensive quantities does not depend on the system volume and is referred to as  an intensive quantity. The studied fluctuation measure $\nu_{dyn}$ is even more complicated because it is inversely proportional to the system volume or to the number of wounded  nucleons within the Wounded Nucleon Model. On the other hand the scaled variance, $(\left<N^2\right>-\left<N\right>^2)/\left<N\right>$, is an intensive quantity. It is therefore more intuitive to get a baseline from p+p interactions for the scaled variance.
The scaled variance of multiplicity distributions, obtained via the \emph{Identity Method} is presented in Fig.~\ref{scaled_var}
for the p+p data of NA61/SHINE (filled circles) and compared to results from 
the EPOS model (solid lines)~\cite{MajaQM}. These are the first released results on fluctuations of identified hadrons in p+p interactions. Therefore, a semi-quantitative agreement with the predictions of the EPOS model is remarkable.  More work is needed to interpret remaining differences between the data and the model.

An additional complication in the experimental study of fluctuations in A+A collisions are unavoidable volume fluctuations from event to event. These additional sources of fluctuations may mask the fluctuations of interest. For example, the scaled variance is sensitive to the volume fluctuations.
A set of so-called \emph{ strongly intensive} quantities, which within a grand canonical ensemble depend neither on the system volume nor on its fluctuations were proposed in Ref~\cite{S_I}. The experimental study of these quantities is ongoing.  

\section{Summary}
We presented preliminary results from NA61/SHINE on inclusive particle production in inelastic p+p collisions 
and discussed its future ion program. Event-by-event fluctuations of particle ratios were analyzed using a novel approach,
the \emph{Identity Method}. Results on excitation functions of $\nu_{dyn}$ in central Pb+Pb collisions confirm an 
increasing trend towards lower energies, reported earlier by NA49. The detailed study of $\nu_{dyn}$ reveals a
strong acceptance dependence at low energies for the $\nu_{dyn}[p,K]$ and $\nu_{dyn}[\pi,K]$. We conclude that the different energy 
dependence of $\nu_{dyn}$ measured by NA49 and STAR ( BES program for central Au+Au collisions) is due to the different phase space 
coverage. Moreover, using the \emph{Identity Method}, the scaled variance of the multiplicity distributions of protons, kaons 
and pions were reconstructed in inelastic p+p collisions of NA61/SHINE. These will serve as an important reference for A+A collisions.

\subsection{Acknowledgments}
The support by the German Research
Foundation (DFG grant GA 1480/2.1) is gratefully acknowledged.

\end{document}